# The Lived User Experience of Virtual Environments: Initial Steps of a Phenomenological Analysis in a Safety Training Setting

**Marko Teräs**  
School of Information Systems  
Curtin University  
Perth, Western Australia  
Email: marko.teras@postgrad.curtin.edu.au

**Hanna Teräs**  
School of Education  
Murdoch University  
Perth, Western Australia  
Email: h.teras@murdoch.edu.au

**Torsten Reiners**  
School of Information Systems  
Curtin University  
Perth, Western Australia  
Email: t.reiners@cbs.curtin.edu.au

## Abstract

Virtual environments (VEs) are making their way into various sectors of life to enhance and support human activity, including learning. VEs have been used in various contexts for training, and in many cases they are designed to model or simulate - as accurately and authentically as possible - a specific work context. In striving for authenticity, visual and representative realism tends to receive most of the development input, despite of several studies that challenge its importance. New training avenues have raised the importance of rigorous phenomenological descriptions for a deeper understanding of user experience in the actual context of use. This paper reports the preliminary steps in a phenomenological analysis of how employees working in actual hazardous settings experience virtual safety training environments. Such open-ended research project can reveal new aspects of user experience that can advice the development and evaluation of human-computer interaction in digital technology-enhanced training contexts.

## 1 Introduction

Recently various forms of virtual environments have received a considerable amount of publicity in business, academic and entertainment publications. Virtual environments enclose a multitude of terms such as virtual reality, virtual worlds, educational games, game-based learning, serious games and simulations. A growing body of literature in safety related training indicates that virtual environments and simulations can develop better spatial awareness and problem solving skills (Tichon and Burgess-Limerick 2011), in addition to engagement and motivation (Reiners et al. 2013). Although an array of research suggests high engagement to be effective in safety training (Burke et al. 2011), disagreements exist about what elements of games and virtual environments actually constitute engagement and immersion. Is it the ever increasing fidelity (Gregory et al. 2013), game elements such as badges, leaderboards and rewards (Deterding et al. 2011), storytelling (Kapp 2012), the affordance to explore, experiment and fail, bringing forth a sense of agency (Freitas and Neumann 2009; Kapp 2012), or something else that previous studies have been unable to fathom? Representational realism often receives majority of the research focus (Wyk and Villiers 2009), while other factors such as the affordance to elicit context, actions, goals and processes have not been studied extensively enough. Studies often note the relationship between context and virtual environments (Dalgarno and Lee 2010; Gamor 2014), but rarely make a deeper effort to understand factors that create a virtual context.

Adequate training and professional development are essential for maintaining and developing professional competence in any profession (Garet et al. 2001). In hazardous environments, it literally becomes a question of life and death. Traditionally, in-company training and professional development have often been "episodic updates of information delivered in a didactic manner, separated from engagement with authentic work experiences" (Webster-Wright 2009, p. 703). Such endeavours typically do not lead to major changes in employees' behaviours and practice: Petraglia (2009) points out that without the ability to bring the information into practice and apply it to relevant contexts, the knowledge lacks authenticity and remains useless.

It is thus not surprising that designers and providers of professional training and development look into possibilities of harnessing virtual environments to achieve more effective and authentic training solutions. This could be highly promising for various sectors such as industry safety training, where more engaging training is reported to relate to more effective knowledge and performance outcomes (Burke et al. 2011). However, the question of what kinds of constituents build the experiential structure of a virtual training environment remains largely unresearched. Concepts such as engagement, immersion and presence are often used in this regard, however, it is often equally unclear what these in fact mean in practice. Moreover, as Petraglia (2009) emphasises, it is the learner's perception that is crucial for learning because learning is always embedded in our own experience of the world, rather than in formal information that can be acquired and memorised. This paper discusses these themes and makes a case for a phenomenological study that sets out to gain an understanding of the professionals' lived user experience of virtual environments in safety training settings.

## 2 Virtual Contexts and Situated Learning

Several authors have discussed the affordances of virtual environments and their ability to mediate a context (Dalgarno and Lee 2010; Gamor 2014). Gee (2008) has suggested that the power of video games is in their ability to embody context-specific knowledge and skills in virtual characters, objects, and environments. They construct an elaborate context-specific (learning) experience for players. Gee (2008) has argued in several cases that all learning is situated. The basis for deeper learning does not lie in the delivery of content and decontextualized facts, but in activity and experience. Thus spaces and actions in games can present learners with situated patterns of play, from which learned knowledge and skills can be transferable to real world settings (de Freitas 2006; McGregor 2007).

Designing learning environments with authentic contexts has been suggested to contribute to deeper learning (Herrington et al. 2010). As Jonassen and Rohrer-Murphy (1999) point out, no human activity can be separated from a context, nor can an activity be analysed outside the context in which it occurs (see more in Dourish 2004). Therefore, instructional design "needs to be more concerned with the context in which learning and performance…occur" (Jonassen and Rohrer-Murphy 1999, p. 62). According to Herrington and her colleagues (2010), an all-embracing authentic context is of importance in providing purpose and motivation for learning. They believe that in designing an authentic context for learning, cognitive realism is more crucial than just physical realism. Authentic context is constructed by knowledge, skills and attitudes used in real settings. Moreover, instead of simplifying the learning context, learning is better facilitated by learning environments that have a realistic level of complexity (Herrington et al. 2010).

The requirement of an authentic context aligns with the idea of learning transfer. Successful transfer occurs when the retrieval conditions are matching with the conditions of learning. In other words, we can better remember and apply what we have learned if the cognitive processes we employ during learning are similar to those that we employ during retrieval (Larsen-Freeman 2013). In other words, if the context is oversimplified or completely different from the context where the information is going to be used, there is a weaker transfer. An example would be a situation where learning takes place in a lecture theatre or seminar room by listening and reading, yet the learner would be required to apply the knowledge in a complex and potentially hazardous work context. This is a typical situation that bases on an assumption of a Cartesian mind-body dualism where mind and external behaviour are separated. Jonassen and Rohrer-Murphy (1999) challenge this assumption and believe that mind and body - mental and physical - are interrelated. Therefore, as they emphasise, "knowing can only be interpreted in the context of doing" (p. 64).

According to Mestre and Vercher (2011), it is important to expose learners to experiences that would be too dangerous to carry out under real conditions, and that virtual environments hold real potential for new knowledge and practical skills acquisition in a safe surrounding (see also de Freitas 2006; McGregor 2007). Dalgarno and Lee (2010) observed that with their realism and interactivity, virtual 3D environments are finally able to model contexts and the different ways skills and knowledge will be used in real life settings. They propose that further research is needed for example in how virtual environment learning contributes to new spatial knowledge, and how greater fidelity and sense of presence lead to improved contextualisation of learning and transfer to real world settings. With modern virtual environments, it is possible to build complex representations of actual work contexts. Still, more research is needed to understand how users experience the use of virtual contexts in specific settings, such as in hazardous environments.

## 3   Phenomenology of Experience

Phenomenology is a systematic study of how phenomena are given to us in consciousness (Giorgi 2012). It aims to access and describe the lifeworld (Lebenswelt) (Husserl 1970), "the world in which we live" (Sokolowski 1999, p. 146). As Cilesiz (2008, p. 240) described, it "is a systematic attempt to come in direct contact with these worlds". Thus it aims to understand the common meaning individuals give to a phenomenon they have lived and experienced (Boland 1986; Creswell 2013; Moustakas 1994). Phenomenology was chosen as the methodological approach of the present study because it can be used to examine the meaning a phenomenon has for individuals. Thus it has the potential to yield rich and unexpected descriptions that would not be found for example with a questionnaire containing predefined questions, even if open-ended ones. As Moustakas (1994, p. 98) points out regarding a phenomenological analysis, "any perspective is a possibility and is permitted to enter into consciousness".

Phenomenology recognizes that accounts that positivistic research calls subjective can give valuable information about the nature of phenomena. It also challenges the Cartesian body-mind dualism: we experience the things themselves, not only our ideas about the things. As Gallagher and Zahavi (2008, p. 6) described:

Husserl's maxim for phenomenology was, 'Back to the things themselves!' (Husserl 1950/1964, p. 6). By this he meant that phenomenology should base its considerations on the way things are experienced rather than by various extraneous concerns which might simply obscure and distort that which is to be understood.

As Sokolowski (1999) has noted, phenomenology aims to reveal matters that are left untouched as too obvious, or which have been cluttered. In the case of VEs, the often taken for granted is for example that what users see, or better graphical realism, correlates with better user experience. Also often certain terms such as 'presence', 'immersion and 'virtual embodiment' are used among professionals to indicate desirable design aims – even if such terms are still ill-defined and debated (see Calleja 2011). Such seemingly innocent preconceptions inevitably advise user experience design. Thus understanding how users experience VEs and what kinds of constituents might give birth to this experience, could contribute to a better design and more meaningful user experience, as well as sharpen the way we discuss VEs.

Boland (1986) argued that phenomenology is a promising way to study information systems: data transforms to information in consciousness, and thus the experience and its structures are what we need to investigate. This study relies on a research tradition that has shown that among other research approaches, phenomenology can contribute important understandings in human-computer interaction (HCI), information systems and computer science (Dourish 2001; Dreyfus 1992; Ihde 1990; Winograd and Flores 1988). Phenomenological analysis of the experience of HCI has already proven to be useful in a wide range of lived experiences such as educational computer use in leisure contexts (Cilesiz 2008), alternative forms of human-data interaction (Hogan 2015), as well as the experience of telepresence in video conferencing (Friesen 2014), to name a few. It has also been suggested to give valuable insights in how users experience new forms of virtual embodiment such as avatars (Ihde 2002; Langdridge 2007).

## 4   Suggested research design and the way forward

Virtual environments and virtual reality are suggested to take us "there" (Heeter 1992), to another environment and context compared to where our physical body is. Popularized descriptions of the virtual make large claims of the potential of virtual environments (Murray 1997; Rheingold 1991), balancing on swaying foundations. New technological developments can create an overly positive and deterministic attitude in using virtual environments and virtual reality in training. We should abstain from making hefty claims that might not be realized, and start from user experience for rigorous understanding of the various uses of VEs. In order to achieve this aim, phenomenology has been chosen to structure the research. The participants of the forthcoming study are working in real life hazardous contexts, and have attended training through a VE. Data will be collected with semi-structured interviews (Bevan 2014; Englander 2012). Instead of theoretical sounding explanations, personal accounts expressed with everyday language will be invited to gather rich accounts on how individuals express meaning of a context in virtual environments (Finlay 2009). The data will be analysed through phenomenological analysis derived from psychology (Giorgi 2009; Moustakas 1994). Common steps in this approach to phenomenological analysis are identifying significant statements and meaning units from the data, rigorously transforming them to discipline-specific language and creating a common meaning structure that describes what constitutes the essence of the experience. As Webster-Wright (2010) describes, "these interlinked commonalities are described in the form of a complex "structure"…that seeks to reveal essential features of that phenomenon".

The study wishes to make a practical contribution for the design of VEs and also for researching and theorizing them. It hopes to result in new knowledge through user accounts that reveal what matters in designing specific contexts with virtual environments in professional training. As it will give a context-specific contribution to the development of VEs for professional safety training, the results might also support designing VEs for students who are not yet working in the industry. Furthermore, as the study highlights common constituents and structures of HCI, its applicability transcends the safety training context and can be useful in other contexts that employ HCI in training. As Ihde (2002, p. 86) has argued, "it is in the interactions, in the mutual questioning and interacting of the world and ourselves, in the changing patters of the lifeworld that things become clear". Thus the study also aims for theoretical significance in discussing the research findings with existing concepts that pursue to explain VE and VR user experience, namely, 'presence', 'immersion', 'virtual embodiment' and 'fidelity'.

# Copyright